\documentclass[aps,pra,preprint,groupedaddress,showpacs]{revtex4}
\usepackage{graphicx}
\usepackage{bm}
\usepackage{floatflt}

\begin{document}

\title{Bound states and scattering lengths of three two-component particles
with zero-range interactions under one-dimensional confinement}
\author{O. I. Kartavtsev}
\author{A. V. Malykh}
\email{maw@theor.jinr.ru}
\affiliation{Joint Institute for Nuclear Research, Dubna, 141980, Russia}
\author{S.~A.~Sofianos}
\affiliation{Physics Department, University of South Africa,
            Pretoria 0003, South Africa}
\date{\today}

\begin{abstract}
The universal three-body dynamics in ultra-cold binary gases confined to
one-dimensional motion are studied.
The three-body binding energies and the (2 + 1)-scattering lengths are
calculated for two identical particles of mass $m$ and a different one of mass
$m_1$, which interactions is described in the low-energy limit by zero-range
potentials.
The critical values of the mass ratio $m/m_1$, at which the three-body states
arise and the (2 + 1)-scattering length equals zero, are determined both for
zero and infinite interaction strength $\lambda_1$ of the identical particles.
A number of exact results are enlisted and asymptotic dependences both for
$m/m_1 \to \infty$ and $\lambda_1 \to -\infty$ are derived.
Combining the numerical and analytical results, a schematic diagram showing
the number of the three-body bound states and the sign of
the (2 + 1)-scattering length in the plane of the mass ratio and
interaction-strength ratio is deduced.
The results provide a description of the homogeneous and mixed phases of atoms
and molecules in dilute binary quantum gases.

\end{abstract}


\maketitle
\section{Introduction}
\label{Introduction}
Dynamics of few particles confined in low dimensions is of interest
in connection with numerous investigations ranging from atoms in ultra-cold
gases~\cite{Gorlitz01,Rychtarik04,Petrov00,Mora04,Mora05,Yurovsky06,Rizzi08}
to nonostructures~\cite{Johnson04,Slachmuylders07,Olendski08}.
Experiments with ultra-cold gases in the one-dimensional (1D) and
quasi-1D traps have been recently
performed~\cite{Gorlitz01,Moritz05,Sadler06,Ospelkaus06}, amid the rapidly
growing interest to the investigation of mixtures of ultra-cold
gases~\cite{Karpiuk05,Shin06,Chevy06,Deh08,Taglieber08,Capponi08,Zollner08}.
Different aspects of the three-body dynamics in 1D have been analyzed in
a number of recent papers, e.~g., the bound-state spectrum of two-component
compound in~\cite{Cornean06}, low-energy three-body recombination
in~\cite{Mehta07}, application of the integral equations in~\cite{Mehta05},
and variants of the hyperradial expansion
in~\cite{Amaya-Tapia98,Amaya-Tapia04,Kartavtsev06}.

It is necessary to emphasize that the exact solutions are known
for an arbitrary number of identical particles in 1D with contact
interactions~\cite{McGuire64,Lieb63}; in particular, it was found
that  the ground-state energy $E_N$ of $N$ attractive particles
scales as $E_N/E_{N=2} = N (N^2 - 1)/6$. There is a vast
literature, in which the exact solution is used to analyze
different properties of few- and many-body systems; few examples
of this approach can be found in
Ref.~\cite{Li03,Girardeau07,Zvonarev07,Guan07}.

The main parameters characterizing the multi-component ultracold
gases, i.~e., the masses and interaction strengths can be easily
tuned within wide ranges in the modern experiments, which handle
with different compounds of ultracold atoms and adjust the
two-body scattering lengths to an arbitrary values by using the
Feshbach-resonance and confinement-resonance
technique~\cite{Olshanii98}. Under properly chosen scales, all the
properties of the system depend on the two dimensionless
parameters, viz., mass ratio and interaction strength ratio, the
most important characteristics being the bound-state energies and
the (2 + 1)-scattering lengths. In particular, knowledge of these
characteristics is essential for description of the concentration
dependence and phase transitions in dilute two-component mixtures
of ultra-cold gases.

In the present paper, the two-component three-body system consisting of a
particle of mass $m_1$ and two identical particles of mass $m$ interacting
via contact ($\delta$-function) inter-particle potential is studied.
In the low-energy limit, the contact potential is a good approximation for any
short-range interaction and its usage provides a universal, i.~e.,
independent of the potential form, description of the dynamics
\cite{Demkov88,Wodkiewicz91,Mehta05,Kartavtsev99,Kartavtsev06,Kartavtsev07}.
More specifically, it is assumed that one particle interacts with the other
two via an attractive contact interaction of strength $\lambda < 0$ while
the sign of the interaction strength $\lambda_1$ for the identical particles
is arbitrary.
This choice of the parameters is conditioned by an intention to consider
a sufficiently rich three-body dynamics since the three-body bound states
exist only if $\lambda < 0$.

Most of the numerical and analytical results can be obtained by solving
a system of hyper-radial equations (HREs)~\cite{Macek68}.
It is of importance that all the terms in HREs are derived analytically;
the method of derivation and the analytical expressions are similar to those
obtained for a number of problems with zero-range
interactions~\cite{Kartavtsev99,Kartavtsev06,Kartavtsev07}.
To describe the dependence on the mass ratio and interaction-strength ratio
for the three-body binding energies and the (2 + 1)-scattering length,
the two limiting cases $\lambda_1 = 0$ and $\lambda_1 \to \infty$ are
considered and the precise critical values of $m/m_1$ for which
the three-body bound states arise and the (2 + 1)-scattering length becomes
zero are determined.
Combining the numerical calculations, exact analytical results, qualitative
considerations, and deduced asymptotic dependencies, one produces a schematic
``phase'' diagram, which shows the number of the three-body bound states
and a sign of the (2 + 1)-scattering lengths in the plane of the parameters
$m/m_1$ and $\lambda_1/|\lambda|$.
This sign is important in studying the stability of mixtures containing both
atoms and two-atomic molecules.

The paper is organized in the following way.
In Sect.~\ref{Outline} the problem is formulated, the relevant notations
are introduced, and the method of "surface" function is described;
the analytical solutions, numerical results and asymptotic dependencies are
presented and discussed in Sect.~\ref{Results}; the conclusions are summarized
in Sect.~\ref{Conclusion}.
\section{General outline and method}
\label{Outline}
The Hamiltonian of three particles confined in 1D, interacting through
the pairwise contact potentials with strengths $\lambda_i$, reads
\begin{equation}
\label{ham}
     H = -\sum_{i} \frac{\hbar^2}{2m_i}\frac{\partial^2}{\partial x_i^2}
     + \sum_{i} \lambda_{i}\delta(x_{jk}) \ ,
\end{equation}
where $x_i$ and $m_i$ are the coordinate and mass of the $i$th particle,
$x_{jk} = x_j - x_k$, and  $ \{ ijk \} $ is a permutation of $ \{ 123 \} $.
In order to study the aforementioned two-component three-body systems, one
assumes that particle 1 interacts with two identical particles 2 and 3 through
attractive potentials and denotes for simplicity $m_2 = m_3 = m$ and
$\lambda_{2} = \lambda_{3} \equiv \lambda<0$.
The corresponding solutions are classified by their parity and are symmetrical
or antisymmetrical under the permutation of identical particles, depending on
whether these particles are bosons or fermions.
The even (odd) parity solutions will be denoted by $P = 0$ ($P = 1$).

In the following, the dependence of the three-body bound state energies and
the (2 + 1)-scattering lengths on two dimensionless parameters $m/m_1$ and
$\lambda_1/|\lambda|$ will be investigated.
Hereafter, one lets $\hbar = |\lambda| = m = 1$ and thus
$m \lambda^2/\hbar^2$ and $\hbar^2/(m |\lambda|)$ are the  units of energy and
length.
Furthermore, one denotes by $A$ and $A_1$ the scattering lengths for
the collision of the third particle off the bound pair of different and
identical particles, respectively.
The scattering length is considered at the lowest two-body threshold, which
corresponds to determination of $A$ if $\lambda_1/|\lambda| >
-\sqrt{2/(1 + m/m_1)}$ and $A_1$ otherwise.
With the chosen units, $E_\mathrm{th} = -1/[2(1 + m/m_1)]$ and
$E'_\mathrm{th} = -\lambda_1^2/4$ are two-body thresholds,~i.e.,
the bound-state energies of two different and two identical particles,
respectively.

The binding energy and the scattering length are monotonic functions of
the interaction's strength and for this reason much attention is paid to
calculations for two limiting cases of zero ($\lambda_1 = 0$) and infinite
($\lambda_1 \to \infty$) interaction between the identical bosons.
It is of interest to recall here that due to one-to-one correspondence of
the solutions~\cite{Girardeau60} all the results derived for systems, in which
the identical particles are bosons and $\lambda_1 \to \infty$, are applicable
to those in which the identical particles are fermions and the s-wave
interaction between them is zero ($\lambda_1 = 0$) by definition.

The numerical and analytical results will be obtained mostly by solving
a system of HREs~\cite{Macek68} where the various terms are derived
analytically~\cite{Kartavtsev99,Kartavtsev06,Kartavtsev07}.
The HREs are written by using the center-of-mass coordinates $\rho$ and
$\alpha$, which are expressed via the scaled Jacobi variables as
$ \rho\sin \alpha =  x_2 - x_3$ and
$\rho\cos\alpha = \cot\omega \left(2 x_1 - x_2 - x_3 \right)$ given
the kinematic-rotation angle $\omega = \arctan\sqrt{1 + 2 m/m_1}$ so that
$E_\mathrm{th} = -\cos^2\omega$.
The total wave function is expanded as in
papers~\cite{Amaya-Tapia98,Amaya-Tapia04,Kartavtsev06,Kartavtsev07},
\begin{equation}
\label{Psi1dim}
       \Psi = \rho^{-1/2} \sum_{n = 1}^{\infty}
                  f_n(\rho)\Phi_n(\alpha , \rho) \,,
\end{equation}
in a set of functions $\Phi_n(\alpha , \rho)$ satisfying the equation at fixed
$\rho$
\begin{equation}
\label{eqPhi}
        \left(\frac{\partial^2}{\partial\alpha^2}
     + \xi^2 \right) \Phi_n(\alpha, \rho) = 0
\end{equation}
complemented by the condition
\begin{equation}
\label{bcomega}
      \frac{\partial\Phi_n(\alpha, \rho)}{\partial\alpha}
           \Bigg|_{\alpha = \omega - 0}^{\alpha = \omega + 0} +
         2\rho\cos\omega\Phi_n(\omega, \rho) = 0 \ ,
\end{equation}
which represents the contact interaction between different
particles~\cite{Wodkiewicz91,Kartavtsev06,Kartavtsev07,Kartavtsev07a}.
Taking into account the symmetry requirements, one can consider the variable
$\alpha $ within the range $0 \leq \alpha \leq \pi/2$ and impose the boundary
conditions
\begin{equation}
\label{boundconda}
    \left[ \left(1 - P \right) \frac{\partial\Phi_n}{\partial\alpha} +
       P \Phi_n \right]_{\alpha = \pi/2} = 0 \,,
\end{equation}
\begin{equation}
\label{boundcondb}
   \left[ \left(1 - T \right) \frac{\partial\Phi_n}{\partial\alpha} +
       T \Phi_n\right]_{\alpha = 0} = 0 \, ,
\end{equation}
where $P = 0$ ($P = 1$) for even (odd) parity and $T = 0$ ($T = 1$) for
$\lambda_1 = 0$ ($\lambda_1 \to \infty$).
These boundary conditions are posed if two identical particles are bosons,
however, the case $T = 1$ is equally applicable if two identical particles are
noninteracting ($\lambda_1 = 0$) fermions.

The solution to Eq.~(\ref{eqPhi}) satisfying the boundary
conditions~(\ref{boundconda}) and ~(\ref{boundcondb})
 can be written as
\begin{equation}
\label{sol1}
      \Phi_n(\alpha, \rho) = B_n
\cases{
\cos [\xi_n (\omega - \pi /2) - P\pi /2)]
     \cos (\xi_n \alpha - T\pi/2)\,, & $ \alpha \le \omega $\cr
  \cos (\xi_n \omega - T\pi/2)
\cos [\xi_n (\alpha - \pi /2) - P\pi /2]\,, & $ \alpha \ge \omega$\cr
}
\end{equation}
where the normalization constant is given by
\begin{eqnarray}
\label{norm1}
       B_n^{2} = -\left[2\cos^2(\xi_n \{\omega - \pi /2\} - P\pi
       /2)
       \cos^2(\xi_n\omega-T\pi/2)\cos\omega\right]^{-1}
                 \frac{{\mathrm d}\xi_n^2}{{\mathrm d} \rho}\ .
\end{eqnarray}
In order to meet the condition~(\ref{bcomega}),
the eigenvalues $\xi_n(\rho)$ should satisfy the equation
\begin{equation}
\label{transeq}
       2 \rho \cos\omega\cos [\xi_n \omega - (\xi_n + P)\pi /2]
         \cos (\xi_n \omega - T\pi/2) + \xi_n
         \sin [(\xi_n + P - T)\pi /2] = 0 \ .
\end{equation}
Notice that the case $P = 1$ and $T = 0$ is formally equivalent
to the case $P = 0$ and $T = 1$ under the substitution of $\omega $
for $\pi /2 - \omega $.

The expansion of the total wave function~(\ref{Psi1dim}) leads to an
infinite set of coupled HREs for the radial functions $f_n(\rho)$
\begin{equation}
\label{system1}
        \left[\frac{{\rm d}^2}{{\rm d} \rho^2} - \frac{\xi_n^2(\rho)
       - 1/4}{\rho^2} + E \right]
       f_n(\rho) - \sum_{m = 1}^{\infty}\left[P_{mn}(\rho) - Q_{mn}(\rho)
       \frac{{\rm d}}{{\rm d}\rho} - \frac{{\rm d}}
        {{\rm d}\rho}Q_{mn}(\rho) \right] f_m(\rho) = 0 \ .
\end{equation}
Using the method described in~\cite{Kartavtsev99,Kartavtsev06,Kartavtsev07},
one can derive analytical expressions for all the terms in
Eq.~(\ref{system1}),
\begin{eqnarray}
\label{Qnm0}
     Q_{nm}(\rho) &\equiv& \langle \Phi_n \bigm|
         \Phi_m'\rangle =\frac{\sqrt{\varepsilon_n'\varepsilon_m'}}
          {\varepsilon_m - \varepsilon_n}\,,
 \\ \label{Pnm0}
     P_{nm}(\rho) &\equiv&
         \langle \Phi_n' \bigm| \Phi_m' \rangle =
\cases{
       \displaystyle Q_{nm}  \displaystyle
       \left[\frac{\varepsilon_n' + \varepsilon_m'}
         {\varepsilon_m - \varepsilon_n} + \frac{1}{2}
       \displaystyle
      \left(\frac{\varepsilon_n''}
           {\varepsilon_n'} - \frac{\varepsilon_m''}
          {\varepsilon_m'}\right)\right]\,,
&      $n\neq m$ \cr
 \displaystyle       - \frac{1}{6}\frac{\varepsilon_n'''}{\varepsilon_n'} +
       \frac{1}{4}\left(\frac{\varepsilon_n''}{\varepsilon_n'}\right)^2\,,
&         $n = m$\cr
}
\end{eqnarray}
where $\varepsilon_n = \xi_n^2$ and the  prime  indicates derivative
with respect to $\rho$.

The obvious boundary conditions for the HREs~(\ref{system1})
$f_n(\rho) \to 0 $ as $\rho \to 0$ and $\rho \to \infty$ was used for
the solution of the eigenvalue problem.
For the calculation of the scattering length $A$, one should impose
the asymptotic boundary condition for the first-channel function
\begin{equation}
\label{f1as}
       f_1(\rho) \sim \rho \sin\omega - A \, ,
\end{equation}
while all other boundary conditions remain the same as for the eigenvalue
problem.
The condition~(\ref{f1as}) follows from asymptotic form of the
threshold-energy wave function at $\rho \to \infty$, which tends to a product
of the two-body bound-state wave function and the function describing
the relative motion of the third particle and the bound pair.
The linear dependence of the latter function at large distance between
the third particle and the bound pair leads to asymptotic
expression~(\ref{f1as}) for the first-channel function in
the expansion~(\ref{Psi1dim}).
On the other hand, the expression~(\ref{f1as}) is consistent with
the asymptotic solution of the first-channel equation in~(\ref{system1}), in
which the long-range terms $P_{11}(\rho)$ and $- 1/(4\rho^2)$ cancel
each other at large $\rho$.
\section{Results}
\label{Results}
%
\subsection{Exact solutions}
\label{Exact}
There are several examples, where the analytical solution of the Schr\"odinger
equation for the systems under consideration can be obtained.
Firstly, for a system containing one heavy and two light particles
(in the limit $m/m_1 \to 0$), using the separation of variables, the solutions
can be straightforwardly written both for zero and infinite interaction
strength between the light particles.
In particular, for $\lambda_1 = 0$, there is a single bound state with
binding energy  $E_{3} = -1$ and the (unnormalized) wave
function is
\begin{equation}
      \Psi_{\rm b} = {\rm e}^{-|x_{12}| - |x_{13}|}\,,
\end{equation}
whereas the scattering wave function at threshold energy
$E_\mathrm{th} = -1/2$ is
\begin{equation}
       \Psi_{\rm sc} = (|x_{12}| - 1)\,{\rm e}^{-|x_{13}|} +
     (|x_{13}| - 1)\, {\rm e}^{-|x_{12}|} \, ,
\end{equation}
which gives the (2 + 1)-scattering length $A = 1$.
On the other hand, for $\lambda_1 \to \infty$, the three-body system is not
bound, and the scattering wave function at the threshold-energy
$E_\mathrm{th} = -1/2$ is
\begin{equation}
        \Psi_{\rm sc} = |x_{12}\, {\rm e}^{-|x_{13}|} - x_{13}\,
{\rm e}^{-|x_{12}|}| \ ,
\end{equation}
which gives $A = 0$.

Furthermore, as mentioned in the introduction, the exact solution is known
for an arbitrary number $N$ of identical particles with a contact interactions
in 1D~\cite{McGuire64,Lieb63} and if the interaction is attractive there is
a single bound state, which energy equals $E_N =- N(N^2 - 1)/24$.
In particular, for three identical particles ($m = m_1$ and
$\lambda_1 = \lambda $) there is  only one bound state with
energy $E_3 = -1$ and the (unnormalized) wave function is
\begin{equation}
   \Psi_{\rm b} =  \exp\left(-\frac{1}{2}\,\sum_{i<j} |x_{ij}|\right)\ ,
\end{equation}
whereas the exact scattering wave function at the two-body threshold
$E_{\rm th} = E'_{\rm th} = -1/4$ is
\begin{equation}
\Psi_{\rm sc} = \sum_{i<j}\exp(- \frac{1}{2}|x_{ij}|) -
4 \exp(-\frac{1}{4}\sum_{i<j}|x_{ij}|)\ ,
\end{equation}
which implies that the (2 + 1)-scattering length is infinite $|A| \to \infty$,
i.~e., there is a virtual state at the two-body
threshold~\cite{Amaya-Tapia98}.

Further exact results can be obtained by using the abovementioned
correspondence of the three-body solutions for the infinite interaction
strength ($\lambda_1 \to \infty$) between two identical bosons and for two
noninteracting fermions ($\lambda_1 \to 0$).
For example, for three equal-mass particles ($m = m_1$) the exact wave
function at the two-body threshold ($E_\mathrm{th} = -1/4$) reads
\begin{equation}
\label{exact1}
     \Psi_{\rm sc} =
\cases{
     {\rm  e}^{-x_{13}/2} + {\rm e}^{x_{12}/2}
        - 2 {\rm e}^{-x_{23}/2}, & $x_{13}\ge 0$  \cr
     |{\rm e}^{x_{13}/2} - {\rm e}^{x_{12}/2}|, & $x_{13}\le 0 $  \, .\cr
}
\end{equation}
As follows from~(\ref{exact1}), the (2 + 1)-scattering length is infinite;
as a matter of fact, this implies a rigorous proof of
the conjecture~\cite{Cornean06} that $m = m_1$ is the exact critical value for
the emergence of the three-body bound state in the case of infinite repulsion
($\lambda_1 \to \infty$) between two identical bosons.

It is worthwhile to recall here the exact solution for three equal-mass
particles ($m = m_1$) if the interaction between two of them is turned off
($\lambda_1 = 0$)~\cite{Gaudin75}.
A transcendental equation was derived for the ground-state energy, which
approximate solution gives the ratio of three-body and two-body energies
$E_3/E_\mathrm{th} \approx 2.08754$.
\subsection{Numerical calculations}
\label{Numerical}
\begin{figure}
\includegraphics[width=8.15cm,height=7.25cm]{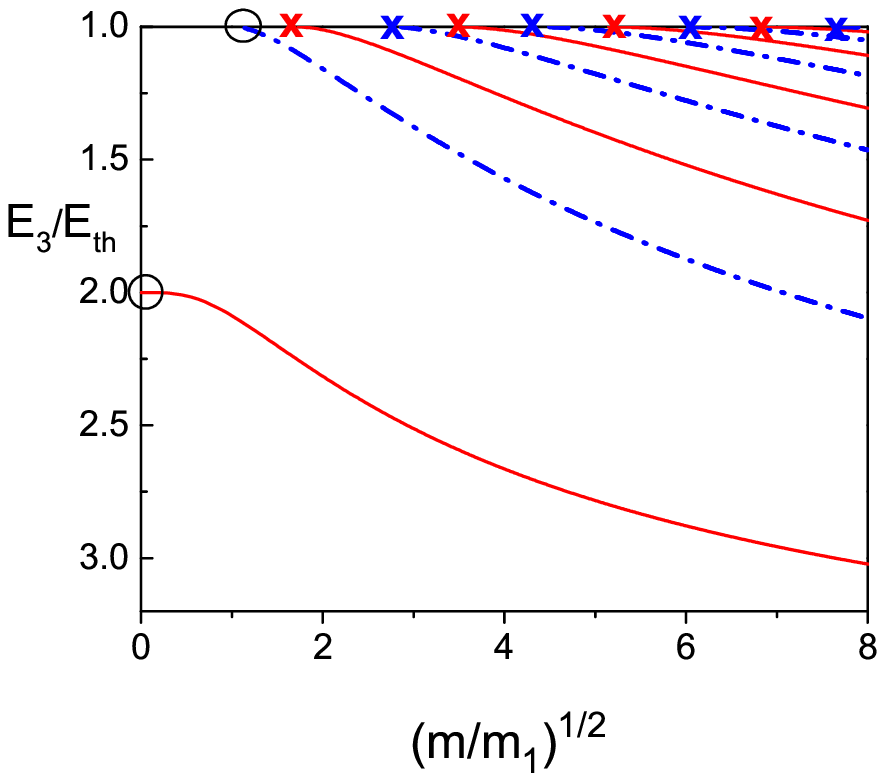}
\includegraphics[width=8.15cm, height=7.25cm]{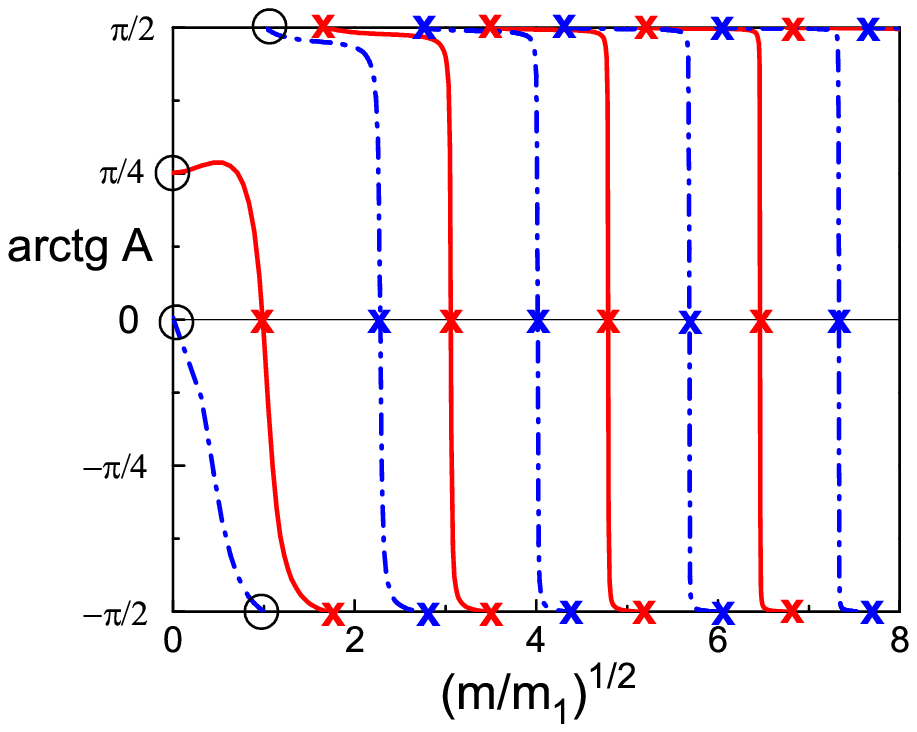}
\caption{\label{fig1}
Mass-ratio dependences for the even-parity states; shown are the ratio of
the three-body bound-state energies to the two-body threshold energy (left)
and the (2 + 1)-scattering length $A$ (right).
Presented are the calculations for a system containing two identical bosons
with zero (solid lines) and infinite (dash-dotted lines) interaction strength
$\lambda_1$.
The dash-dotted lines represent also the results for a system containing two
identical noninteracting ($\lambda_1 = 0$) fermions.
Encircled are those points, in which the exact analytical solution is known.}
\end{figure}

For the even-parity states ($P = 0$) and the two limiting values of
the interaction strength between identical bosons, $\lambda_1 = 0$ and
$\lambda_1 \to \infty$, the HREs~(\ref{system1}) are solved to determine
the mass-ratio dependence of three-body binding energies and
the (2 + 1)-scattering length $A$.
The calculations show sufficiently fast convergence with increasing the number
of channels; 15-channel results are presented in Fig.~\ref{fig1}.
%
The precise critical values of the mass ratio, for which the three-body bound
states arise ($|A| \to \infty$) and the (2 + 1)-scattering length $A = 0$ are
presented in Table~\ref{table1} and are marked by crosses in Fig.~\ref{fig1}
and Fig.~\ref{fig3}.
%
\begin{table}[htb]
\caption{The even-parity critical values of the mass ratio $m/m_1$ for which
the (2 + 1)-scattering length becomes zero (marked by $A = 0$) and an $n$th
three-body bound state arises (marked by $|A| \to \infty$).
Calculations done for two values of the interaction strength between
the identical particles, $\lambda_1 = 0$ and $\lambda_1 \to \infty$.
\label{table1} }
\begin{tabular}{ccc|cc}
\hline\hline
   & \multicolumn{2}{c|}{$\lambda_1 = 0$} &
\multicolumn{2}{c}{ $\lambda_1 \to \infty$}\\
\hline
$n$ & $m/m_1 (A = 0)$ & $m/m_1 (|A| \to \infty)$ & $m/m_1 (A = 0)$ &
$m/m_1 (|A| \to \infty)$  \\
\hline
 1 &  -       &  -        &   0$^*$  &  $1^*$   \\
 2 &  0.971   &  2.86954  &  5.2107  &  7.3791  \\
 3 &  9.365   &  11.9510  &  16.1197 & 19.0289  \\
 4 &  22.951  &  26.218   &  32.298  & 35.879   \\
 5 &  41.762  &  45.673   &  53.709  & 57.923   \\
 6 &  65.791  &  70.317   &  80.339  & 85.159   \\
 7 &  95.032  &  100.151  &  112.179 & 117.583  \\
 8 &  129.477 &  135.170  &  149.222 & 155.193  \\
 9 &  169.120 &  175.374  &  191.463 & 197.989  \\
10 &  213.964 &  220.765  &  238.904 & 245.973  \\
\hline  \hline
$^*$ Exact
\end{tabular}
\end{table}
%
The condition that the ground state energy is twice the threshold energy is
important as it determines whether production of the triatomic molecules is
possible in a gas of diatomic molecules.
The mass ratio, at which $E_3/E_\mathrm{th} = 2$ is determined to be
$m/m_1 \approx 49.8335$ for $\lambda_1 \to \infty$, while for the excited
states the condition $E_3/E_\mathrm{th} = 2$ is satisfied for
$m/m_1 \approx 130.4516 $ if $\lambda_1 = 0$ and $m/m_1 \approx 266.1805 $
if $\lambda_1 \to \infty$.

As shown in Fig.~\ref{fig1}, the binding energies increase with increasing
the mass ratio, whereas, the scattering length $A$ has a general trend to
decrease with increasing the mass ratio on each interval between two
consecutive critical mass ratios at which the bound states appear.
Nevertheless, the calculations for $\lambda_1 = 0$ show that $A(m/m_1)$ becomes
non-monotonic function at small $m/m_1$.
More precisely, the scattering length takes a maximum value $A \approx 1.124$
at $m/m_1 \approx 0.246$.
Again one has to note that the mass-ratio dependence of energy and scattering
length (plotted in Fig.~\ref{fig1}) and the critical values of the mass ratio
(presented in Table~\ref{table1}) are the same both for the three-body system
containing two identical bosons if $\lambda_1 \to \infty$ and for
the three-body system containing two identical noninteracting
($\lambda_1 = 0$) fermions.

It is of interest to note that the calculated binding energy
$E_3/E_\mathrm{th} \approx 2.087719$ for three equal-mass particles
($m = m_1$) if two identical ones do not interact with each other
($\lambda_1 = 0$) is very close to the result~\cite{Gaudin75}
$E_3/E_\mathrm{th} \approx 2.08754$ obtained from the analytical
transcendental equation (see Sect.~\ref{Exact}).
A small discrepancy most probably stems from the approximations
of~\cite{Gaudin75} made in numerical solution of the transcendental equation.
The (2 + 1)-scattering length turns out to be small and negative,
$A \approx -0.09567 $, for $m = m_1$ and $\lambda_1 = 0$ and takes a zero
value at slightly smaller mass ratio $m/m_1 \approx 0.971$ (see
Table~\ref{table1}).

\begin{figure}
\includegraphics[width=.55\textwidth]{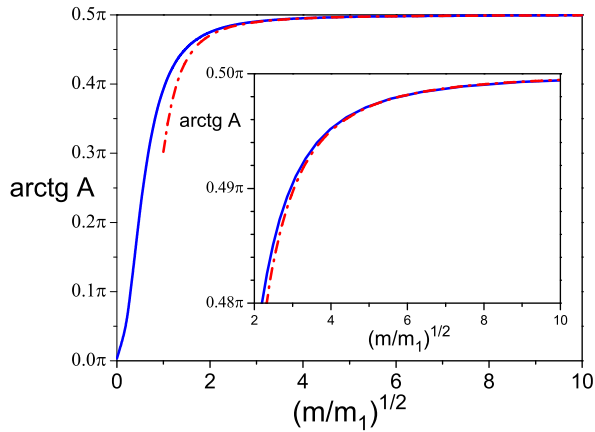}
\caption{\label{fig2}
Mass-ratio dependence of the (2 + 1)-scattering length $A$ for odd-parity
states ($P = 1$) of a system containing two identical noninteracting bosons
($\lambda_1 = 0$).
The numerical calculation (solid lines) is compared with the large-mass-ratio
asymptotic behaviour given by Eq.~(\ref{sclP1asymp}) (dash-dotted lines).
The dependence corresponding to large $A > 15$ is shown on a large scale in
the inset. }
\end{figure}

Analogously, the odd-parity (P = 1) solutions for three-body system containing
two identical noninteracting bosons ($\lambda_1 = 0$) were obtained.
As follows from Eq.~(\ref{transeq}), the eigenvalues $\xi_n(\rho)$ entering in
HREs~(\ref{system1}) are nonnegative, which implies that there are no
three-body bound states.
The calculated dependence of the scattering length $A$ is shown in
Fig.~\ref{fig2}; $A$ increases monotonically with increasing mass ratio
following the asymptotic dependence discussed in Sect~\ref{Qualitative}.
%
\subsection{Asymptotic dependencies}
\label{Qualitative}
%
\subsubsection{Large attractive interaction of two identical particles }
\label{Linter}
In the limit of large attractive interaction between the identical
particles, $\lambda_1 \to - \infty$, the even-parity wave function
takes, with a good accuracy, the factorized form $\Psi \simeq
\phi_0(x_{23}) u(y)$ [$y =\cot\omega \left(2 x_1 - x_2 - x_3
\right)$], where $\phi_0(x) = \sqrt{|\lambda_1|/2}\exp(-|\lambda_1
x|/2)$ is the wave function of the tightly bound pair of identical
particles with energy $E'_{\rm th} = -\lambda_1^2/4$ and $u(y)$
describes the relative motion of a different particle 1 with
respect to this pair. Within this approximation, $u(y)$ is a
solution of the equation
\begin{equation}
\label{eqrel}
        \left[\frac{{\rm d}^2}{{\rm d} y^2} +
       2|\lambda_1| \exp{(- \sqrt{1 + 2m/m_1}|\lambda_1y|)} +
       \lambda_1^2/4 + E \right] u(y) = 0 \, ,
\end{equation}
which gives the independent of $\lambda_1$ leading-order terms in
the asymptotic expansion for the three-body binding energy,
$\varepsilon \approx 4/(1 + 2m/m_1)$, and the (2 + 1)-scattering
length,
\begin{equation}
\label{A1as}
          A_1 \approx (1 + 2m/m_1)/4 \, .
\end{equation}
\subsubsection{Two heavy and one light particles}
\label{Asymptotic}
For large mass ratio $m/m_1$, one can use the adiabatic and quasi-classical
approximations which provide, e.~g., a universal description
for the energy spectrum~\cite{Kartavtsev07a}.
To describe the three-body properties in the limit of large $m/m_1 \to \infty$
[$\omega \to \pi/2 - \sqrt{m_1/(2m)}$], one considers the first eigenvalue
$\xi_1(\rho) \equiv i\kappa(\rho)$, which large-$\rho$ asymptotic dependence
is approximately given by
\begin{equation}
\label{eigeneqminfty}
         \rho \cos\omega = \frac{\kappa}{1 + (-1)^P
             {\rm e}^{-\kappa(\pi - 2\omega)}} \ ,
\end{equation}
as follows from Eq.~(\ref{transeq}) on the equal footing for the system
containing two identical bosons both for $\lambda_1 = 0$ and
$\lambda_1 = \infty$ and for the system containing two identical
noninteracting fermions.

The number of the three-body even-parity ($P = 0$) bound states $n$ can be
determined for large $m/m_1$, using the one-channel
approximation in~(\ref{system1}) and the effective potential
$-\kappa^2(\rho)/\rho^2$,  from~(\ref{eigeneqminfty}).
Within the framework of the quasi-classical approximations and taking into
account the large-$\rho$ asymptotic dependence~(\ref{eigeneqminfty}), one
obtains the relation $m/m_1 \approx C (n + \delta)^2$ in the limit of large
$n$ and $m/m_1$.
The constant $C$ can be found as
\begin{equation}
\label{Cquasicl}
       C = \frac{\pi^2}{2}\left[\int_0^1\sqrt{2t + t^2}
      \frac{1 + (1 - \ln t)t}{2t(1 + t)^2}dt\right]^{-2}\approx 2.59 \ ,
\end{equation}
where the integral is expressed by letting $t = \exp[-\kappa(\pi - 2\omega )]$
in the leading term of the quasi-classical estimate,
\begin{equation}
\label{intminfty}
       \cos\omega\int_0^\infty {\mathrm d}\rho \left\{ \left[(1 +
        {\rm e}^{\displaystyle\kappa(\rho)\left(\pi - 2\omega
        \right)}\right]^2 - 1\right\} ^{1/2} = \pi n \ .
\end{equation}
Fitting the calculated mass-ratio dependence of the critical values, at which
the bound states appear, to the $n$-dependence $C (n + \delta)^2$ (up to
$n = 20$, see Table~\ref{table1} for 10 lowest values), one obtains in a good
agreement with the quasi-classical estimate~(\ref{Cquasicl}) $C \approx 2.60$
both for $\lambda_1 \to \infty$ and $\lambda_1 = 0$.
Simultaneously, one obtains $\delta = 0.73$ if $\lambda_1 \to \infty$ and
$\delta = 0.22$ if $\lambda_1 = 0$ for the parameter, which determine
the next-to-leading order term of the large-$n$ expansion.

The asymptotic dependence of the effective potential
$-\kappa^2(\rho)/\rho^2$ obtained from~(\ref{eigeneqminfty})
allows one to find the leading order mass-ratio dependence of the
odd-parity ($P = 1$) scattering length,
\begin{equation}
\label{sclP1asymp}
        A = \frac{m}{m_1} \sqrt{1 + \frac{m_1}{2m}}
        \left(\ln\frac{m}{m_1} + 2 \gamma \right) \ ,
\end{equation}
where $\gamma \approx 0.5772$ is the Euler constant. The
convergence of the calculated dependence $A(m/m_1)$ to the
asymptotic dependence~(\ref{sclP1asymp}) is shown in
Fig.~\ref{fig2} for the case of two identical noninteracting
bosons ($\lambda_1 = 0$).
\subsection{Mass-ratio and interaction-strength ratio dependencies}
\label{dependencies}
\begin{figure}[thb]
\includegraphics[width=0.7\textwidth]{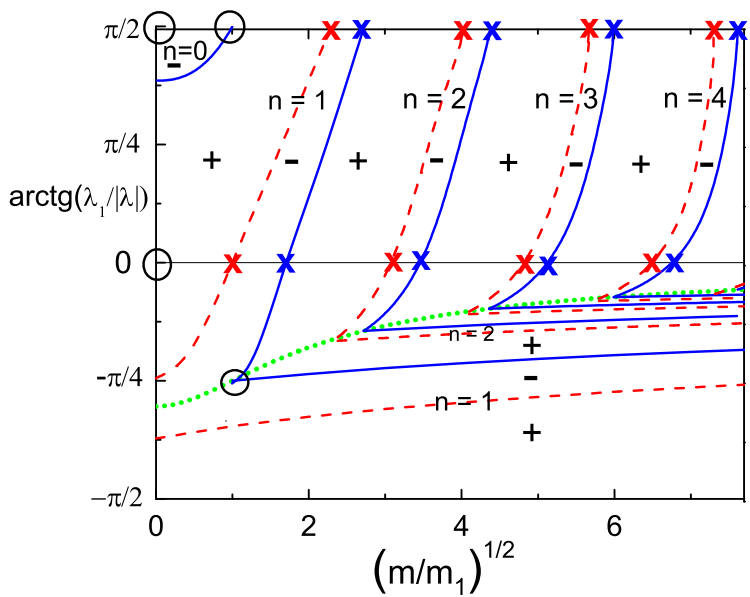}
\caption{\label{fig3}
Schematic ``phase'' diagram for the even-parity states of two identical bosons
and the third different particle.
The dotted line marks the border between two areas where the lowest two-body
threshold is set by the energy of two different and two identical particles.
The number of the three-body bound states is marked by $n$ in
the corresponding areas separated by solid lines.
The sign of the (2 + 1)-scattering lengths $A$ and $A_1$ is marked by $\pm$
and the corresponding areas are separated by dashed lines.
The crosses show the calculated critical values of the mass ratio (enlisted in
Table~\ref{table1}).
Encircled are those points, in which the exact analytical solution is known.}
\end{figure}

Collecting the numerical and the exact analytical results, the asymptotic
expressions, and qualitative arguments, one obtains a schematic ``phase''
diagram, which depicts the number of three-body bound states and the sign
of the (2 + 1)-scattering lengths in the $m/m_1$ - $\lambda_1/|\lambda|$ plane
(shown in Fig.~\ref{fig3}).

%
%
The plane of parameters is divided into two parts by a dotted line,
$\lambda_1/|\lambda| = -\sqrt{2/(1 + m/m_1)}$, with the low-energy three-body
properties being essentially different in the upper and lower part, where
the two-body threshold is determined by the bound-state energy of two
different and identical particles, respectively.
The lines which represent the condition $|A|= \infty$ or $|A_1| = \infty$
(arising of the three-body bound state) separate areas with different number
of the bound states, whereas the conditions $A = 0$ or $A_1=0$ split each area
into two parts of different signs of the scattering lengths.

It can be proven rigorously that in the upper part of the diagram (above
the dotted line), the number of the three-body bound states $n$ increases and
the (2 + 1)-scattering length $A$ decreases with decreasing the interaction
strength $\lambda_1$, while in the lower part (below the dotted line) $n$
increases and $A_1$ decreases with decreasing the mass ratio $m/m_1$.
The proof is based on the representation for which the lowest two-body
threshold is independent of $\lambda_1$ and $m_1$ in the former and latter
case, respectively.
The required conclusion follows from the monotonic dependence of
the Hamiltonian on $\lambda_1$ and $m_1$.
A schematic ``phase'' diagram demonstrated in Fig.~\ref{fig3}, is drawn by
using more strict assumption on the positive slope of the lines, which
show where the three-body bound states arise ($|A| \to \infty$) and where
the (2 + 1)-scattering lengths ($A = 0$ and $A_1 = 0$ in the upper and lower
part of the $\lambda_1/|\lambda|$ - $m/m_1$ plane, respectively) become zero.
Tentatively, this assumptions seems to reflect correctly the general trend;
nevertheless, one should note that the slope of the isolines of constant
scattering length is not generally positive.
In particular, $A$ is not a monotonic function of the mass ratio for
$\lambda_1 = 0$, as shown in Fig.~\ref{fig2}; this implies a non-monotonic
dependence of the constant-$A$ isolines in a region near the point
($m/m_1 = 0$, $\lambda_1/|\lambda| = 0$).

For sufficiently large repulsion $\lambda_1$ and small mass ratio $m/m_1$
the three-body bound states are lacking.
The limit $m/m_1 \to 0$ (1D analogue of the helium atom with contact
interactions between particles) was discussed in paper~\cite{Rosenthal71},
where the binding energy as a function of the repulsion strength between light
particles was calculated and the critical value of the repulsion strength for
which the three particles becomes unbound was determined.
Recently, a very precise critical value $\lambda_1/|\lambda| \approx 2.66735$
was found in~\cite{Cornean06}.
The boundary of the $n = 0$ area (shown in the upper left corner in
Fig.~\ref{fig3}) goes from the point ($m/m_1 = 0$,
$\lambda_1/|\lambda| \approx 2.66735$) to the point ($m/m_1 = 1$,
$\lambda_1 \to \infty$), as was conjectured in~\cite{Cornean06} and proven in
Sect.~\ref{Exact} by using the exact solution at the latter point.
Taking into account this result, the above-discussed monotonic dependence on
$\lambda_1$, and the exact solution for three identical particles, one comes
to an interesting conclusion that there is exactly one bound state ($n = 1$)
of three equal-mass particles independently of the interaction strength
$\lambda_1$.
There is exactly one bound state ($n = 1$) also for a sufficiently large
attraction between identical particles whereas the second bound state
appears for $m > m_1$ and $ |\lambda_1| < 1$ (as shown in Fig.~\ref{fig3}).
Therefore, the scattering length $A_1$ changes from the positive value given
by~(\ref{A1as}) at $\lambda_1 \to - \infty $ to the negative one
as $\lambda_1$ increases.
The strip areas corresponding to $n > 1$ are located at higher values of
the mass ratio with the large-$n$ asymptotic dependence
$n \propto \sqrt{m/m_1}$.
In each parameter area corresponding to $n$ bound states, the scattering
lengths run all the real values tending to infinity at the boundary with
the $n - 1$ area and to minus infinity at the boundary with the $n + 1$ area.

\section{Conclusion}
\label{Conclusion}
%
The three-body dynamics of ultra-cold binary gases confined to
one-dimensional motion is studied. In the low-energy limit, the
description is universal, i.~e., independent of the details of the
short-range two-body interactions, which can be taken as a sum of
contact $\delta$-function potentials. Thus, the three-body
energies and the (2 + 1)-scattering lengths are expressed as
universal functions of two parameters, the mass ratio $m/m_1$ and
the interaction-strength ratio $\lambda_1/|\lambda|$. The
mass-ratio dependences of the binding energies and the scattering
length are numerically calculated for even and odd parity and the
accurate critical values of the mass ratio, for which the bound
states arise and the scattering length became zero, are
determined. It is rigorously proven that $m/m_1 = 1$ is the exact
boundary, above which at least one bound state exists (as
conjectured by~\cite{Cornean06}); the related conclusion is the
existence of exactly one bound state for three equal-mass
particles independently of the interaction strength between the
identical particles. Asymptotic dependences of the bound-state
number and the scattering length $A$ in the limit $m/m_1 \to
\infty$ and of the binding energy and the scattering length $A_1$
in the limit $\lambda_1 \to -\infty$ are determined. Combining the
numerical calculations, analytical results, and qualitative
considerations, a schematic diagram is drawn, which shows the
number of the three-body bound states and the sign of the (2 +
1)-scattering length as a function of the mass ratio and
interaction-strength ratio.

The obtained qualitative and quantitative results on the three-body properties
provide a firm base for description of the equation of state and phase
separation in dilute binary mixtures of ultra-cold gases.
In particular, a sign of the (2 + 1)-scattering lengths essentially controls
the transition between the homogeneous and mixed phases of atoms and diatomic
molecules.
The condition $E_3/E_\mathrm{th} > 2$ defines the parameter area, where
the production of the triatomic molecules is energetically favorable in a gas
of diatomic molecules.

From the analysis of the ``phase'' diagram in Fig.~\ref{fig3} it follows that
still there are interesting problems deserving further elucidation.
These include the problem of non-monotone dependence of the constant-$A$
isolines in the $\lambda_1/|\lambda|$ - $m/m_1$ plane, the behaviour of
the lines separating the positive and negative scattering lengths within
the $n = 1$ area, and the description of the beak formed by the lines
separating the $n = 1$ and $n = 2$ areas in the vicinity of the exact solution
for three identical particles ($\lambda_1 = \lambda$ and $m = m_1$).

One should discuss the connection of the present results with those,
which take into account the finite interaction radius $R_e$ and (quasi)-1D
geometry.
The determination of the corrections due to finite interaction radius is
not a trivial task, however, one expects that the corrections should  be small
for all calculated values provided $R_e/a$ and $R_e/a_1$ are small, where
$a$ and $a_1$ are the two-body scattering lengths.
On the other hand for sufficiently tight transverse confinement, one
expects that the main ingredient is the relation between the 3D and quasi-1D
two-body scattering lengths established in~\cite{Olshanii98}.
Moreover, a role of the transverse confinement does not simply reduce to
renormalization of the scattering lengths; the full scale three-body
calculations are needed to determine the energy spectrum and the scattering
data in the (quasi)-1D geometry.

It is worthwhile to mention that more few-body problems are of interest in
binary mixtures.
In particular, the low-energy three-body recombination plays an important role
in the kinetic processes, while the elastic and inelastic cross sections for
collisions either of diatomic molecules or of atoms off triatomic molecules
are needed to describe the properties of the molecular compounds.
\section*{Acknowledgements}
This work is based upon research supported by the National Research Foundation
(NRF) of South Africa within the collaborating agreement between
the Department of Science and Technology of South Africa and the Joint
Institute for Nuclear Research, Russia.

\bibliography{onedim}

\end{document}